\documentclass[aps,prd,twocolumn,superscriptaddress,amsfonts,amssymb,amsmath,eqsecnum,nofootinbib,floatfix,showpacs]{revtex4}

\usepackage{graphicx,color,ulem}
\definecolor{darkblue}{rgb}{0,0,0.7}
\definecolor{darkred}{rgb}{0.7,0,0}
\usepackage[unicode, colorlinks, citecolor=darkblue, linkcolor=darkred, urlcolor=blue]{hyperref}

\allowdisplaybreaks

\begin{document}
\date{\today}
\title{Thermal noise of beam splitters in laser gravitational wave detectors}
\author{Johannes~Dickmann}
\email{johannes.dickmann@ptb.de}
\affiliation{Physikalisch-Technische Bundesanstalt, Bundesallee 100, 38116 Braunschweig, Germany}

\author{Stefanie~Kroker}
\affiliation{Physikalisch-Technische Bundesanstalt, Bundesallee 100, 38116 Braunschweig, Germany}
\affiliation{Technische Universit\"at Braunschweig, LENA Laboratory for Emerging Nanometrology, Pockelsstra{\ss}e 14, 38106 Braunschweig, Germany}

\author{Yuri~Levin}
\affiliation{Department of Physics, Columbia University, 704 Pupin Hall, 538 West 120th st, New York, NY 10027, USA}
\affiliation{Center for Computational Astrophysics, Flatiron Institute, 162 5th Ave, New York, NY 10010, USA}
\affiliation{School of Physics and Astronomy, Monash University, Clayton, VIC 3800, Australia}

\author{Ronny~Nawrodt}
\affiliation{5. Physikalisches Institut, Universit\"at Stuttgart, Pfaffenwaldring 57, 70550 Stuttgart, Germany}

\author{Sergey~Vyatchanin}
\affiliation{Faculty of Physics, M.V. Lomonosov Moscow State University, Moscow 119991, Russia}
\affiliation{Quantum Technology Centre, M.V. Lomonosov Moscow State University, Moscow, Russia}

\begin{abstract}
We present the calculation of thermal noise in interferometric gravitational-wave detectors due to the thermal fluctuations of the beam splitter (BS). This work makes use of a recently developed method of the analysis of thermal noise in mirrors from first principles, based on the fluctuation dissipation theorem. The evaluation of BS thermal noise is carried out for the two different gravitational wave observatories, GEO600 and the Advanced Laser Interferometer Gravitational Wave Observatory (aLIGO). The analysis evaluates thermal noise from both the substrate and the optical reflective and antireflective stacks located on the BS surface. We demonstrate that the fluctuations of both reflecting and anti-reflecting surfaces significantly contribute to the total thermal noise of the BS. The oscillating intensity pattern couples small-scale distortions of the surface to the overall phase readout, and therefore increases the overall thermal noise. In the case of aLIGO, the BS contribution is with 0.3\% negligibly small. At a frequency of 500\,Hz, the BS causes about 10\% of GEO600's sensitivity limit. BS noise impairs the feasible sensitivity of the GEO-HF design proposal by about 50\%.

\end{abstract}


\maketitle

%

\section{Introduction}

The first direct detections of gravitational waves by the advanced LIGO detectors \cite{GW_2016, GW_2016b, GW_2017, GW_2017b} have opened the era of gravitational wave astronomy. These fascinating detection results are a consequence of technology breakthroughs allowing us to measure tiny displacements of $\sim 10^{-18}$\,m of macroscopic test masses \cite{aLIGO2013,aLIGO2014,aLIGO2015,aLIGO2015b, aVirgo2015, grote2010}. It is the thermal noise of the test masses that sets a severe limitation for the detector sensitivity in their most sensitive frequency band from $50$~Hz to $2000$~Hz \cite{99a1BrLeVy,99a1BrGoVy,00a1BrGoVy,HarryBook2012}. In current detectors, a major source for thermal noise is Brownian structural noise \cite{nawrodt2011} in mirror coatings. The thermally induced random stresses in the coatings and the substrates of the test masses produce random deformations of their surfaces, which are detected as thermal noise at the interferometer output.

A gravitational wave induces a variation of the frequency of the main mode in the arm cavity of the interferometer, which is registered as a phase shift of a monochromatic optical wave reflected from the cavity. Concurrently, the thermal noise of the mirror's surface also randomly changes the eigenfrequency of the mode masking the gravitational wave signal. 
The same considerations hold true for the thermal noise of the beam splitter (BS) in the interferometer. Usually, the beam splitter is a cylindrical plate made of an optically transparent material. One surface is covered by a reflecting (R) coating and the other by an anti-reflecting (AR) one. The thermal fluctuations of \textit{both} surfaces also change the interferometer's eigenfrequency. In previous works, BS noise was estimated by the approximation of BS as infinitely thin plates \cite{harmsPRD2004} and in the simplifications of small light beam radii, compared to the BS size and light at normal incidence \cite{somiya2009b}.

In this paper, we present the accurate computation of BS thermal noise in gravitational wave detectors from first principles, following the approach formulated in \cite{18a1TuLeVyPLA}. 
In combination with the fluctuation-dissipation theorem (FDT) \cite{Callen1951PR, LL5,Levin1998} this approach allows the accurate computation of the thermal noise resulting from thermal fluctuations of both R and AR surfaces of BS with light beams of finite size and oblique incidence. A similar approach has been used in \cite{17a1KrDiHeNaLeVy,18plaDiHuNaKr} for computing thermal noise in reflective gratings and in \cite{15a1DeGo} for evaluating the influence of an absorbing layer on the resonant frequencies and Q-factors of spherical microresonators.

Here we apply our approach for the calculation of Brownian noise in substrate and coating of the BS for arbitrary polarizations of the light. Our estimates show that the contributions of thermoelastic \cite{99a1BrGoVy} and thermorefractive \cite{00a1BrGoVy, benthem2009} noise are subdominant to the Brownian noise computed here for frequencies between 100 and 4000 Hz.

\begin{figure}[b]
\includegraphics[width=0.45\textwidth]{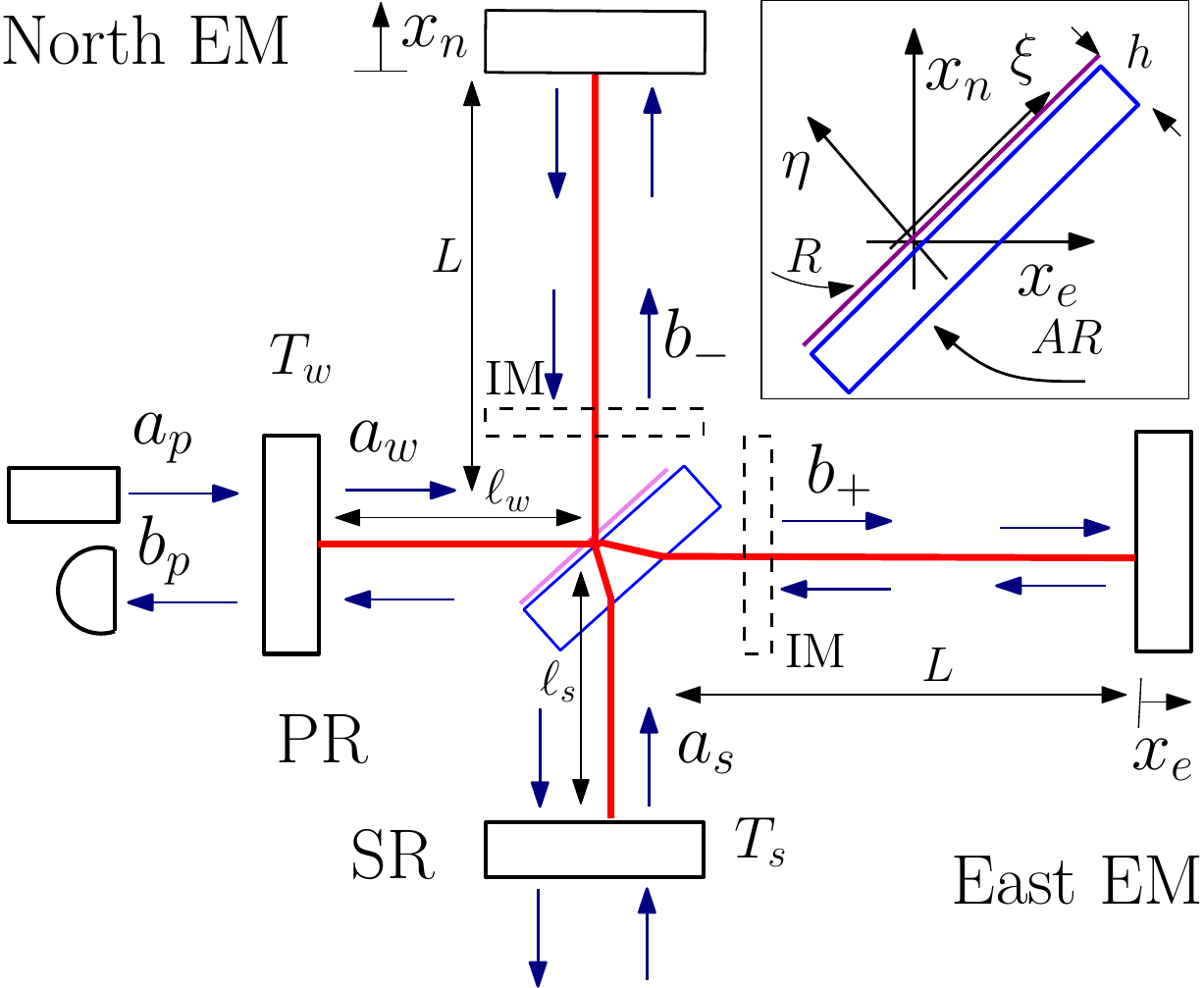}
\caption{Schematic view of the BS plate in GW interferometers. Fluctuations of R and AR surfaces (see inset) produce noise in the detecting dark port. The end mirrors are denoted by EM. SR and PR are signal and power recycling mirrors with the intensity transmissions $T_w$ and $T_s$, respectively. In the case of the advanced LIGO interferometers, additional input mirrors (IM) in the arms (shown by dashed lines) are illustrated. In GEO600 these additional mirrors are absent.}\label{DCsimple}
\end{figure}

\section{Model and statement of the problem}

We consider the simplified model of the GEO600 interferometer as shown in Fig.~\ref{DCsimple}. We assume that the length of north (east) arm differs from a multiple of the laser wavelength $\lambda$ by small displacements $x_n$ ($x_e$):
\begin{align}
 L_n &= n\lambda +x_n, \quad L_e= \ell\lambda +x_e,\quad L_n\simeq L_e\simeq L,
\end{align}
where $n$ and $\ell$ are integer numbers. We also assume that the length $L$ of north and east arm is much larger than the length $\ell_w$ of west arm and length $\ell_s$ of the south arm (see notations in Fig.~\ref{DCsimple}):
\begin{align}
   \label{Lell}
 L \gg \ell_s,\ \ell_w.
\end{align}

To begin with, we assume that the power recycling (PR) and signal recycling (SR) mirrors are perfectly reflecting. In the case of perfectly tuned arms, we have two optical modes: the ``west" mode (the standing wave is in the \textit{west} arm and is absent in the south arm) and the ``south" mode (the standing wave is in the \textit{south} arm and absent in the west arm). The displacements $x_e,\ x_n$ of the end mirrors produce a coupling between the two modes. Thus, the Hamiltonian $\mathcal H$ for the two modes is written in the following form (for details see \cite{Law1995}):
\begin{align}
    \mathcal H &= \hslash \omega_w a_w^*a_w\left(1-\frac{x_+}{L}\right) +
		\hslash \omega_s a_s^*a_s \left(1-\frac{x_+}{L}\right)\nonumber\\
    \label{Hcross}
	  	 &\quad   + \hslash \sqrt{\omega_s \omega_w}\big(a^*_wa_s+a_wa^*_s\big)\frac{x_-}{L},\\
  \label{xpm}
	  	 &\boxed{x_\pm \equiv \frac{x_e\pm x_n}{2}.}
\end{align}
The cross term of the Hamiltonian is responsible for the occurrence of a gravitational wave signal at the photo detector. It provides information on small variations of the differential coordinate $x_-$. 
Deformations of the BS surfaces also result in a coupling of the two modes. Therefore, we must calculate the coupling coefficients and particularly the cross term to translate the BS fluctuations into effective fluctuations $x_-^\text{eff}$ of the differential coordinate. 

Let us consider the eigenfrequencies $\omega_w$, $\omega_s$ and $\omega_0$ \eqref{Hcross} for this particular case: 
\begin{align}
\label{omega0}
\omega_w=\omega_s,\ \omega_0 = \omega_s \left(1 - \frac{x_+}{L}\right).
\end{align}
Instead of the partial field coordinates $a_s,\ a_w$ for two coupled modes (oscillators), we introduce 
eigen (normal) coordinates $b_\pm$ and rewrite the Hamiltonian as follows:
\begin{subequations}
 \label{H2}
  \begin{align}
   \mathcal H &= \mathcal H_+ + \mathcal H_-,\quad \mathcal H_\pm \equiv \hslash \omega_\pm b_\pm^* b_\pm ,\\
   \label{bpm}
    & b_\pm \equiv \frac{a_w \pm a_s}{\sqrt 2},\quad 
	 \omega_\pm = \omega_0\left(1 \pm \frac{x_-}{L}\right).
  \end{align}
 \end{subequations}
 The $b_+$ mode in the east arm is called "east" mode (it is absent in the north arm) and the $b_-$ mode is called "north" mode. In the relations \eqref{H2}, there are two independent oscillators and for each of them we calculate the adiabatic invariants $\mathcal I_\pm$ and the resulting frequency variations and variations of the energies $\mathcal E_\pm$:
 \begin{subequations}
 \label{modespm}
 \begin{align}
  \mathcal I_+ &= \frac{\mathcal E_+}{ \omega_+},\quad
   \mathcal I_- = \frac{\mathcal E_-}{ \omega_-},\\
   \label{Deltaomega}
     \frac{\Delta \omega_+}{\omega_+} & = \frac{\Delta \mathcal E_+}{\mathcal E_+}  , \quad
      \frac{\Delta \omega_-}{\omega_-} = \frac{\Delta \mathcal E_-}{\mathcal E_-}.
 \end{align}
The adiabatic invariant of an oscillator is conserved, if its frequency changes very slowly compared to its oscillation period (here: the period of the optical oscillations). We are now interested in the effective small changes of $x_-$ and $x_+$, created by small perturbations of the BS surface. In our case, variations of the eigenfrequencies above can be written using (\ref{omega0}, \ref{bpm}): 
  \begin{align}
  \label{setpm}
    \frac{\Delta \omega_+}{\omega_+} & = -\frac{x_+}{L} + \frac{x_-}{L},\quad 
    \frac{\Delta \omega_-}{\omega_-}  = -\frac{x_+}{L} - \frac{x_-}{L},
  \end{align}
By using the Eqs. \eqref{setpm} and \eqref{Deltaomega}, we find the effective relative displacements:
  \begin{align}
      \label{Epm}
  - \frac{x_+}{L} &=  \frac{\Delta \mathcal E_+}{\mathcal E_+} + \frac{\Delta \mathcal E_-}{\mathcal E_-}\,,\\
  \label{xminus2}
   \frac{x_-}{L} & =  \frac{\Delta \mathcal E_+}{\mathcal E_+} - \frac{\Delta \mathcal E_-}{\mathcal E_-}\, .
  \end{align}
\end{subequations}
Let us assume that a small perturbation of the BS surface appears slowly on the initially flat surface. This will change $\Delta \mathcal E_+$ and $\ \Delta \mathcal E_-$, which can be computed from the elastic energy stored in the BS due to the applied ponderomotive light pressures. As the surface perturbations are small, we can apply the presented approximations and calculate the pressures with Maxwell stress tensor of the \textit{unperturbed} field in the region around the surface deformation. 

\subsection{Calculations of the effective relative displacement $(x_-/L)$}

\begin{figure}[b]
\includegraphics[width=0.25\textwidth]{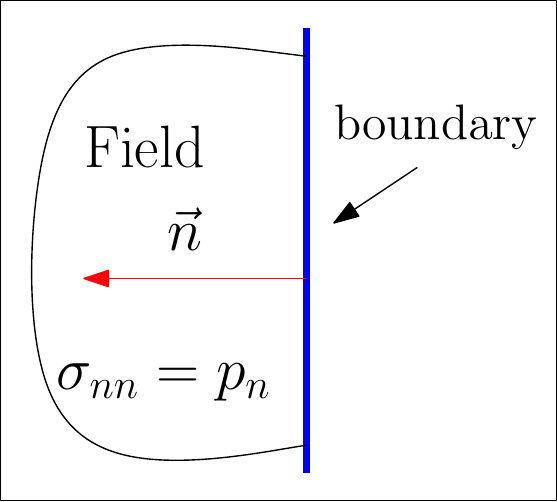}
\caption{Maxwell stress tensor $\sigma_{nn}$ leads to pressure $p_n$ along \textit{outer} normal $\vec n$ to the boundary.}\label{Mstress}
\end{figure}

For the calculation of the terms in \eqref{xminus2} it is convenient to use the eigenamplitudes $b_\pm$ \eqref{bpm}. The work performed against the pressure is equal to the change of energy in the cavity taken with opposite sign. The light pressure $p_i$ acting on dielectric media's surface may be calculated by means of Maxwell stress tensor $\sigma_{ij}$ inside and outside the material:
\begin{align}
\label{sigmaD}
 \sigma_{ij}^\epsilon &= \frac{1}{4\pi}\left(\epsilon E_iE_j +H_iH_j -\frac{\epsilon E^2+H^2}{2}\,\delta_{ij}\right).
\end{align}
The pressure in \eqref{sigmaD} is directed along the \textit{outer} normal of the surface. For example, for a reflecting surface of the BS, the stress tensor is equal to the pressure acting along $(\eta)-$axis (see Fig.~\ref{DCsimple}). In contrast, the stress tensor calculated for the fields directly \textit{beneath} the reflecting surface (inside the BS) is equal to the pressure acting along the $(-\eta)-$axis (compare Fig.~\ref{Mstress}).

The total normal pressure is equal to the difference of the pressures inside and outside the BS at the surface:
\begin{align}
\label{pi}
  p_i=p_i^\text{outer} - p_i^\text{inside}.
\end{align}
The relative energy variations $\Delta \mathcal E_\pm $ can be calculated as the work of ponderomotive light forces resulting from the total normal pressure:
\begin{align}
\label{DeltaE}
 \frac{\Delta \mathcal E_\pm }{\mathcal E_\pm} &= - \int F_\pm (\vec r_\bot)\,u(\vec r_\bot)\, d\vec r_\bot,\quad
  F_\pm = \frac{p_\pm (\vec r_\bot)}{\mathcal E_\pm },
 \end{align}
where $ u(\vec r_\bot)$ is a small perturbation of the BS surface in normal direction, depending on the coordinate $\vec
r_\bot$ on the BS surfaces. The averaging functions $F_\pm$ are defined by the pressures $p_\pm$ calculated for the 
corresponding modes and have to be applied to \textit{both} sides of the BS, as will be shown later. For small 
perturbations $\Delta \mathcal E \ll\mathcal E$ the functions $F_\pm$ depend only on geometric factors, i.e. surface 
perturbations, radius of light beam, refractive index and thickness of the BS. They do not depend on the mode amplitudes 
$b_\pm$, because the energies $\mathcal E_\pm$ and the pressures $p_\pm$ are both proportional to $\sim |b_\pm|^2$.

\subsection{Brownian Noise Power Spectral density}\label{SD}

Following the FDT \cite{Callen1951PR, LL5,Levin1998},  we have to apply virtual pressures $p(\vec r) =F_0 p_\pm(\vec r)\sin\omega t$, oscillating with an angular frequency $\omega$, to calculate the mean dissipated power $W(\omega)$ in the BS and to finally calculate the Brownian noise power spectral density $S_{BS}(\omega)$:
\begin{align}\label{SBS}
 S_{BS}(\omega) &= \frac{8 k_BT W(\omega)}{\omega^2 F_0^2},
\end{align}
\noindent where $k_B$ is the Boltzmann constant and $T$ is the absolute temperature.
In this paper, we are interested in structural Brownian noise \cite{Saulson1990PRD}. In the model of structural loss the dissipated power $W$ is defined by the phenomenological loss angle $\phi$:
\begin{align}
\label{Energy}
 W = \mathbb E \omega\phi,
\end{align}
where $\mathbb E$ is the mean elastic energy stored in the BS.
The pressures on the beam splitter $p_\pm$ have striped patterns \cite{14a1HeCrGrHiLuNaSiVaVyWi,18a1TuLeVyPLA} i.e. which are proportional to $\sim \cos^2 k\xi/\sqrt 2$ ($k\equiv \omega_0/c$, $c$ is the speed of light). 
Hence, the applied pressure can be divided into two parts: 1) a smooth (non-striped) pressure $P_\text{sm}$ and 2) a striped  pressure $P_\text{str}\sim \cos \sqrt 2 k\xi$. The elastic energies for the problems 1 and 2 can be calculated separately. Indeed, the total energy can be calculated as integral over both sides of the BS:
\begin{subequations}
\begin{align}
 \mathbb E&=\frac{1}{2}\int \left(P_\text{sm}(\vec r_\bot)+P_\text{str}(\vec r_\bot)\right)\\
	&\qquad \times	\left(u_\text{sm}(\vec r_\bot)+u_\text{str}(\vec r_\bot)\right)d\vec r_\bot\nonumber\\
	&= \mathbb E_\text{sm} + \mathbb E_\text{str} + \mathbb E_\times ,\\
 \mathbb E_\text{sm}	&= \frac{1}{2}\int P_\text{sm}(\vec r_\bot)u_\text{sm}(\vec r_\bot)\, d\vec r_\bot,\\
 \mathbb E_\text{str}	&=	 \frac{1}{2}\int P_\text{str}(\vec r_\bot)u_\text{str}(\vec r_\bot)\, d\vec r_\bot\\
 \mathbb E_\times &=\frac{1}{2}\int \left[ P_\text{sm}(\vec r_\bot)u_\text{str}(\vec r_\bot
	  + P_\text{str}(\vec r_\bot)u_\text{sm}(\vec r_\bot))\right] d\vec r_\bot\,,
\end{align}
\end{subequations}
where $u_\text{sm}(\vec r_\bot)$ and $u_\text{str}(\vec r_\bot)$ are surface displacements caused by smooth and striped pressures, respectively. Obviously, the cross term of energy $\mathbb E_\times$ will be negligibly small after integration over the surface, due to the fast oscillating multiplier $\cos \sqrt 2 k\xi$.
To our best knowledge, there is no approach to solve problem 1 analytically. Hence,
 we solve it numerically using the finite element tool COMSOL Multiphysics \cite{COMSOL}. In contrast, problem 2 can be solved analytically using the well-known solution for a half infinite elastic media \cite{14a1HeCrGrHiLuNaSiVaVyWi}. This method can be applied, because the pressure contribution with fast spatial oscillation, characterized by the wave vector $k_0$, leads only to deformations located close to the surface. In particular, the elasticity divergence $\Theta=u_{xx} + u_{yy} + u_{zz}$ decreases along the $z$-axis normal to the surface as $\sim e^{-kz}$.

\section{Calculations of pressures}

We assume all light beams in the interferometer to have an amplitude Gaussian distribution over the cross section:
  \begin{align}
 \label{f0}
 f_0(r_\bot) = \frac{e^{-r_\bot^2/2r_0^2}}{\sqrt{\pi r_0^2}},\quad \int|f_0|^2 \, d\vec r_\bot =1,
\end{align}
where $d\vec r_\bot \equiv r_\bot \, dr_\bot\, d\phi$. For gravitational wave detectors we can assume that the beam radius $r_0$ is \textit{large}: 
\begin{align}
\label{condr0}
  r_0\gg \lambda=\frac{2\pi}{k},
\end{align}
where $\lambda$ is the wavelength. Hence, we can consider the wave in the cavity as a plane wave with an amplitude multiplied by the Gaussian factor $f_0$ omitting terms of higher orders $\sim 1/kr_0$ (see for example \cite{kogelnik1966, davis1979, davis1981}). For the calculations, we introduce the following relations between the coordinates $(x_e,\ x_n)$ and $(\xi,\ \eta)$ (see Fig.~\ref{DCsimple}):
\begin{align}
 x_e &=\frac{\xi -\eta}{\sqrt 2},\quad x_n = \frac{\xi +\eta}{\sqrt 2},\\
 \xi &= \frac{x_e + x_n}{\sqrt 2},\quad \eta =\frac{- x_e + x_n}{\sqrt 2}\,.
\end{align}
Inside the BS the light propagates with an angle $\alpha$ with respect to the BS axis (see Fig.~\ref{BSplusS}):
\begin{align}
  \label{sin}
 \sin \alpha =\frac{1}{n\sqrt 2} ,\quad  2a= h \tan \alpha.
\end{align}
On the AR surface there are two centers of Gaussian distributions separated by distance $2a$ (see Fig.~\ref{BSplusS}). 
Inside the BS, the beams propagate along the axes $y_e,\ y_n$. These coordinates may be recalculated from the 
coordinates $(\xi,\, \eta)$:
\begin{align}
 y_e & =\xi \sin \alpha + \eta \cos\alpha,\\
 y_n &= \xi\sin\alpha - \eta \cos \alpha. 
\end{align}
We calculate the fields and pressures for different polarizations of traveling waves: S-polarization (vector of electric field is normal to the plane of figure) and p-polarization (vector of magnetic field is normal to the plane of figure). As a result, after simple but cumbersome calculations presented in Appendix \ref{app1}, we obtain the averaging functions on both surfaces of the BS and for each polarization orientation. We present these functions in the following 2 subsections.

\subsection{S-polarization summary}\label{secS}

For the GEO600 interferometer and s-polarization, we obtain the following equation for the eigenfrequency shift, i.e. effective relative displacement $x_-/L$. For the reflecting surface R of the BS we retrieve using \eqref{xminus2} and (\ref{prS+}, \ref{prSmb})
\begin{align}
  \left.\frac{x_-}{L}\right|_{rS} &=\int \frac{F_r^s(\vec r_\bot)}{L}\,u_\bot(\vec r_\bot)\, d\vec r_\bot,\\
  \frac{F_r^s(\vec r_\bot)}{L} &\equiv  \left(\frac{p^{rS+}}{\mathcal E_+} - \frac{p^{rS-}}{\mathcal E_-}\right)\\
    \label{Rsmode}
  & =    \frac{f_\bot^2}{2L}\left\{\left[2n -\frac{1}{n} +1\right]\right.\\
  &\qquad  \left. -\left[1 + \frac{1}{n} +2\sqrt 2(n-1) \right] \cos 2 k_{bs}\xi\right\}, \nonumber
  \end{align}
  where we use the notations of \eqref{sin} and
  \begin{align}
     \label{kbs}
     k_{bs} & = \frac{k}{\sqrt 2},\\
     \label{fbot}
     f_\bot &\equiv \frac{1}{\sqrt{\pi r_0^2}} 
      \exp\left(-\frac{z^2 + 0.5\xi^2}{2r_0^2}\right).
  \end{align}  
  For the anti-reflecting surfaces AR of the BS we get with \eqref{xminus2} and (\ref{parS+E}, \ref{parS-}):
  \begin{align}
   \left.\frac{x_-}{L}\right|_{arS} &=\int \frac{F_{ar}^s(\vec r_\bot)}{L}\,u_\bot(\vec r_\bot)\, d\vec r_\bot, \\ 
  \frac{F_{ar}^s(\vec r_\bot)}{L} &\equiv  \left(\frac{p^{arS+}}{\mathcal E_+} - \frac{p^{arS-}}{\mathcal E_-}\right)\\
    \label{ARsmode}
   &=  \frac{1}{2L}
  \left\{  \left(   f_+^2\right)\left[1 - 2n +\frac{1}{n}\right]\right.\\
  &\quad \left.
   + 2\sqrt 2(n-1)f_- f_+ \cos\kappa(\xi_+ + \xi_-)   \right.\nonumber\\
  &\qquad \left.
      -   f_+^2  \left(1 - \frac{1}{n}\right) 
	 \cos  2\kappa \xi_+  \right\}.\nonumber
\end{align}
  Here, we use the following notations (\ref{sin}, \ref{kbs}, \ref{fbot}) and definitions:
  \begin{align}
   \label{fpm}
 f_{\pm} & = \sqrt\frac{1}{\pi r_0^2}\exp\left(-\frac{z^2+0.5(\xi\mp a)^2}{2r_0^2}\right),\\
 & \label{xipm} \xi_\pm \equiv \xi \mp a\,.
  \end{align}

\subsection{P-polarization summary}\label{secP}

For the GEO600 interferometer and p-polarization, we obtain the following equations for the eigenfrequency shift, i.e. effective $x_-/L$. For the surfaces R we get using \eqref{xminus2} and (\ref{FrP+}, \ref{prPm})
\begin{align}
  \left.\frac{x_-}{L}\right|_{rP} &=\int \frac{F_{r}^p(\vec r_\bot)}{L}\,u_\bot(\vec r_\bot)\, d\vec r_\bot,\\
    \frac{F_{r}^p(\vec r_\bot)}{L} &\equiv  \left(\frac{p^{rP+}}{\mathcal E_+} - \frac{p^{rP-}}{\mathcal E_-}\right)\\
    \label{Rpmode}
   &=
  \frac{f_\bot^2}{2L}\left\{\left[2n - \frac{1}{n} +1\right]  \right.\\
  &\quad \left.+\left[1+ \frac{1}{n}  + 2\sqrt 2\big(n-1\big)\right] 
  \cos\sqrt 2 k\xi\right\}. \nonumber
\end{align}
For the surface AR we get using \eqref{xminus2} and (\ref{parP+}, \ref{parP-})
\begin{align}
  \left.\frac{x_-}{L}\right|_{arP}& =\int \frac{F_{ar}^p(\vec r_\bot)}{L}\,u_\bot(\vec r_\bot)\, d\vec r_\bot, \\ 
  \frac{F_{ar}^p(\vec r_\bot)}{L} &\equiv  \left(\frac{p^{arP+}}{\mathcal E_+} - \frac{p^{arP-}}{\mathcal E_-}\right)\\
  \label{ARpmode}
    &=  \frac{1}{2L}
  \left\{  \left(   f_+^2\right)\left[1 - 2n + \frac{1}{n}\right]\right.\\
  &\qquad \left. - 2\sqrt 2(n-1)f_- f_+ \cos \kappa(\xi_+ +\xi_-)  \right.\nonumber\\
  &\qquad \left. +\left(  f_+^2\right)
      \left(1 - \frac{1}{n}\right)  \cos   2k_{bs}\xi_+  \right\}. \nonumber
\end{align}
We see that the "smooth" parts of the pressure acting on R and AR surfaces are the same both for p- and s-polarizations (compare \eqref{Rsmode} with \eqref{Rpmode} and \eqref{ARsmode} with \eqref{ARpmode}). However, the 
"striped" contributions have opposite signs for s- and p-polarization while being equal in their absolute value. Hence, for non-polarized light and for the case of equal s- and p-polarized field amplitudes, the ``striped" term vanishes.

\subsection{Generalization for the aLIGO interferometers}

The results in subsections \ref{secS} and \ref{secP} are obtained for the GEO600 configuration  \textit{without} input mirrors (IM), shown by dashed rectangles in Fig.~\ref{DCsimple}.  For the calculation of the mode energies $\mathcal E_\pm$, we use the fact, that the amplitudes of the fields on BS are nearly the same as in the arms, and in addition we utilize the approximation \eqref{Lell}. For advanced LIGO, we have to recalculate the energies $\mathcal E_\pm$ by taking into account, that the mean amplitudes in the arms are by a factor of approximately $2/\sqrt T_\mathrm{IM}$ larger than on the BS. Here $T_\mathrm{IM}$ represents the power transmittance of the IM. Hence, we can generalize, for example, formula \eqref{Rsmode} for s-polarization on the R surface by the transformation:
\begin{align}
\label{rule}
 F_r^s|_\text{GEO} &=  F_r^s|_\text{aLIGO}\times \frac{T_\mathrm{IM}}{4}.
\end{align}
Obviously, the other formulas (\ref{ARsmode}, \ref{Rpmode}, \ref{ARpmode}) can be generalized for advanced LIGO using the same transformation \eqref{rule}.

\begin{table}[b]
   \caption{Parameters of BS for GEO600 and aLIGO.}\label{table1}
	 \begin{tabular}{|p{0.25\textwidth} | r|r|}
	 \hline \hline
	 Parameters & aLIGO & GEO600   \\
	 \hline \hline
	 Radius $R$ of BS, m & $0.1875$ & $0.13$ \\
	 Height $h$ of BS, m & $0.064$  & $0.08$\\
	 Refractive index $n_{\text{SiO}_2}$  & $1.45$ & $1.45$\\
   $	a =\frac{h}{\sqrt{2n^2 -1}}$ & $0.036$ & $0.045$\\
	 Radius  $w_0=\sqrt 2 r_0$ of light beam on BS, m 
	 \footnote{$r_0$ is the beam radius for the \textit{intensity} distribution, which is proportional to $\sim \exp[-r^2/r_0^2]$, whereas $w_0$ is the radius for the \textit{amplitude} distribution, which  is proportional to $\sim \exp[-r^2/w_0^2]$). So  $w_0 = \sqrt 2\, r_0$.} & $ 0.06$ & $0.0088$\\
	 Young's modulus $E_{\text{SiO}_2}$, GPa & 73.1 & 73.1\\
	 Poisson's ratio $\nu_{\text{SiO}_2}$ &0.17 &0.17 \\
	 Density $\rho_{\text{SiO}_2}$, kg/m$^3$ & 2203 & 2203\\ 
	 Loss angle $\phi_{\text{SiO}_2}$ & $10^{-8}$  & $10^{-8}$\\
	 Power transmittance of input mirrors in arms of aLIGO & $5\times 10^{-3}$ & --- \\
	 \hline \hline
	 \end{tabular}
\end{table}

\section{Calculation of the elastic energy}

To calculate the spectral density of the Brownian BS noise in terms of relative displacements $x_-/L$ for s-polarization, we have to apply virtual pressures $p_{rs}$ to the R surface of the BS using \eqref{Rsmode} and $p_{ars}$ to the AR surface using \eqref{ARsmode}:
\begin{align}\label{press}
 p_{rs}= F_0 F_r^s(\vec r) \sin\omega t,\quad p_{ars}= F_0 F_{ar}^s(\vec r) \sin\omega t,
\end{align}
where $F_0$ is a constant, see \eqref{SBS}. For the p-polarization we should use (\ref{Rpmode}, \ref{ARpmode}). Then we calculate the mean elastic energy $\mathbb E$ stored in the BS and substitute this result into \eqref{SBS} by taking \eqref{Energy} into account. 
Since the elastic problem cannot be solved analytically, we performe the computations numerically using COMSOL \cite{COMSOL}. We use the parameters for the BS of GEO600 and advanced LIGO listed in Table~\ref{table1}. For the following considerations, we account only for the smooth contributions of the pressures \eqref{press}, which are equal for both polarizations (see (\ref{Rsmode}, \ref{ARsmode}, \ref{Rpmode}, \ref{ARpmode}). For the solution of the elastic problem, we have to fulfill two conditions: a) the sum of all external forces and b) the total torque of all external forces should be equal to zero. However, the pressures integrated over the R and the AR surfaces are not equal to zero, i.e. the total force acting on the BS is not equal to zero. Hence, in analogy to inertial forces we have to apply an additional volume force $f$, which is \textit{homogeneously distributed} over the BS such that the total force is equal to zero \cite{Liu2000}:
\begin{align}
 f &= - \frac{1}{\pi R^2 h}\int \left[  p_{r}(\vec r_\bot) + p_{ar}(\vec r_\bot)\right] d S=\\
   &= -\frac{\sqrt 2 F_0}{\pi R^2h}\times (1-\epsilon_F) ,
\end{align}
where the integration is performed over the area of the BS, having radius $R$ and height $h$. In approximation $R\to \infty$, the coefficient $\epsilon_F$ is zero. For the parameters of advanced LIGO and GEO600 in Table \ref{table1}, the numerical computations yield:
\begin{align}
 \epsilon_F^\text{LIGO} &\simeq -0.000142,\\
  \epsilon_F^\text{GEO} &\simeq 3.3\times 10^{-16}.
\end{align}
The total torque of the external forces is not zero, because the centre of $p_{ar}$ is shifted from the symmetrical cylinder axis by distance $a$. Hence, we have to introduce an additional volume force $f_\text{add}\sim y/R$ in order to eliminate effective torques:
\begin{align}
 f_\text{add} &= (1-\epsilon_T) \frac{\sqrt 2\, F_0}{\pi R^2 h}\left(1 - 2n +\frac{1}{n} \right)
      \times \frac{2a y}{R^2},
\end{align}
where the small coefficient $\epsilon_T$ results from the finite dimension of the BS. In the approximation $R\to \infty$, we retrieve $\epsilon_T\to 0$. For the parameters in Table~\ref{table1}, numerical computations lead to: 
\begin{align}
 \epsilon_T^\text{LIGO} &\simeq 0.00576962,\\
  \epsilon_T^\text{GEO} &\simeq 1.19\times 10^{-13}.
\end{align}

\subsubsection{Substrate thermal Brownian noise of the BS} 

In order to calculate Brownian noise from thermal fluctuation in the BS substrate, we calculate the mean elastic energy stored in the BS. Using COMSOL, we performed numerical calculations of the mean elastic energy for parameters in Table~\ref{table1}: 
\begin{align}
   \label{Eligo}
		\mathbb E_\text{aLIGO} &= 3.98 \times 10^{-10} \ J\times \frac{F_0^2}{N^2}, \\
   \label{Egeo}
		\mathbb E_\text{GEO} &= 1.97 \times 10^{-9} \ J\times \frac{F_0^2}{N^2}.
\end{align}
Using \eqref{SBS}, we compute the power spectral density $S_\text{BS-}$ of the BS Brownian noise, recalculated to the differential coordinate $x_-$:
\begin{align}
   \label{estLIGO}
 \left.\frac{\sqrt{S_\text{BS-}(\omega)}}{L}\right|_\text{LIGO}
   &\simeq 4.5\times 10^{-27}\, \frac{1}{\sqrt\text{Hz}}, \\
   \label{estGEO}
 \left.\frac{\sqrt{S_\text{BS-}(\omega)}}{L}\right|_\text{GEO}
   &\simeq 2.7\times 10^{-23}\, \frac{1}{\sqrt\text{Hz}},
\end{align}
at the angular frequency $\omega= 2\pi\times 100 \, \text{s}^{-1} $ for advanced LIGO  and GEO600, respectively. For advanced LIGO we have taken the transformation \eqref{rule} into account. The current sensitivity of advanced LIGO \cite{aLIGO2015b} is about $\sqrt{S_x}/L \simeq 5\times 10^{-23}\, 1/\sqrt\text{Hz}$, for the future cryogenic LIGO Voyager \cite{Voyager2014} the planned sensitivity is about $\sqrt{S_x}/L \simeq 8\times 10^{-25}\, 1/\sqrt\text{Hz}$. Since the substrate BS noise is substantially smaller than this sensitivity, LIGO Voyager will not be limited by it. The current sensitivity of GEO600 \cite{grote2010} is about $\sqrt{S_x}/L \simeq 3\times 10^{-22}\, 1/\sqrt\text{Hz}$. It is about 10 times larger than BS noise evaluated here.

\begin{table}[b]
   \caption{Parameters of the reflecting and anti-reflecting coatings of the BS for advanced LIGO \cite{coyne2005,  billingsley2018}.}\label{table2}
	 \begin{tabular}{|p{0.28\textwidth} | r|r|}
	 \hline \hline
	 Parameters & SiO$_2$ & Ta$_2$O$_5 $  \\
	 \hline \hline
	 Refractive index & 1.45 & 2.1\\
	 Young modulus, GPa & 73.1 & 140\\
	 Poisson ratio  &0.17 &0.23 \\
	 Loss angle & $1\times 10^{-4}$  & $4\times 10^{-4}$\\
	 R coating total thickness $h_r$, nm & $ 523$ &  $320$\\
	 AR coating total thickness $h_{ar}$, nm & $517$ & $ 359$\\
	 \hline \hline
	 \end{tabular}
\end{table}

\subsubsection{Coating thermal Brownian noise of the BS}

\begin{figure}
\includegraphics[width=0.5\textwidth]{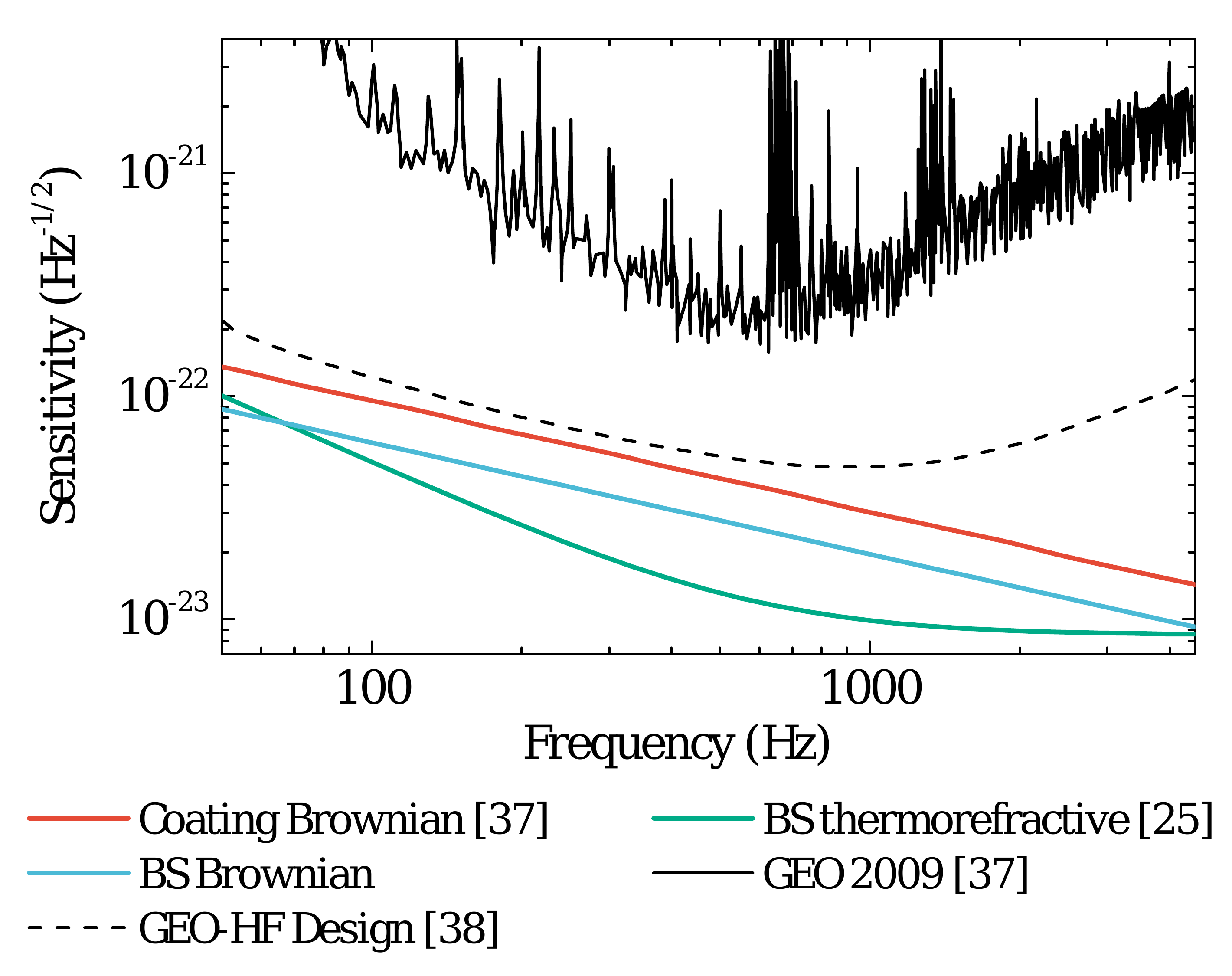}
\caption{Noise $\sqrt{S}/L$ of GEO600 in $1/\sqrt{\text{Hz}}$ versus mechanical readout frequency $f$ for the following thermal noise contributions: The mirror coating Brownian \cite{Lueck2010JOP}, the BS thermorefractive \cite{benthem2009} and the BS Brownian noise computed in this work. Additionally, the measured sensitivity of 2009 \cite{Lueck2010JOP} and the design sensitivity of GEO-HF \cite{Grote2010CQG} are shown.}\label{fig:GEO600sensitivity}
\end{figure}

The Brownian noise of the R and AR coatings has to be calculated separately, because the loss angle of the alternating layers of $Ta_2O_5$ and $SiO_2$ are much larger than the loss of the substrate. These coatings are required to provide the optical function of the reflective and anti-reflective BS surfaces. For advanced LIGO, the parameters of the BS coatings are listed in Table~\ref{table2}. The parameters of the GEO600 coatings are not published. Therefore, we assume the optical coating design to be the same as for LIGO (see Table~\ref{table2}).
We calculate the energy stored in the coatings with the assumption that the coatings are thin compared to the BS thickness. Thus, the strains in the layers are approximately the same as on the upper side of the substrate. We use the following expression for the volume density $w$ of elastic energy in the layer \cite{Saulson1990PRD, 14a1HeCrGrHiLuNaSiVaVyWi}:
\begin{align}
\label{wn2}
  w_n & = \frac{1}{2} \frac{(1+\nu)(1-2\nu) \sigma_{zz}^2}{Y(1-\nu)}  \nonumber\\
	 &\quad + \frac{Y \,(u_{xx}+u_{yy})^2}{4(1-\nu)}+ \frac{Y \,(u_{xx}-u_{yy})^2}{4(1+\nu)} \\
	 &\qquad + \frac{Y}{1+\nu}\times u_{xy}^2. \nonumber
\end{align}
Here $\sigma_{zz}$ is the normal component of the stress tensor, $Y$ and $\nu$ are Young's modulus and Poisson's ratio of the layers (made of Ta$_2$O$_5$ or SiO$_2$), whereas $u_{ij}$ are the tangent components of the strain tensors, which are the same as for the substrate's surface. The total energy stored in the SiO$_2$ and Ta$_2$O$_5$ layers of the R and the AR coatings can be calculated as:
 \begin{align}
  \mathbb E_\text{coat}&= h_{r\, \text{SiO}_2} \int w_n|_{r\, \text{SiO}_2}\, dS_r\\
    &\qquad +h_{r\, \text{Ta}_2\text{O}_5} \int w_n|_{r\, \text{Ta}_2\text{O}_5}\, dS_r\\
    &\qquad + h_{ar\, \text{SiO}_2} \int w_n|_{ar\, \text{SiO}_2}\, dS_r \\
    &\qquad +h_{ar\, \text{Ta}_2\text{O}_5} \int w_n|_{ar\, \text{Ta}_2\text{O}_5}\, dS_r,
 \end{align}
where the integration is carried out over the R and AR surfaces. 
Performing a numeric integration using $u_{ij}$, obtained for the substrate and the parameters listed in Table~\ref{table2}, we obtain at $\omega=2\pi 100$ s$^{-1}$:
\begin{align}
   \label{est2GEO}
 \left.\frac{\sqrt{S_\text{BS-}(\omega)}}{L}\right|_\text{GEO}
   &\simeq 6.2\times 10^{-23}\, \frac{1}{\sqrt\text{Hz}},\\
   \label{est2LIGO}
 \left.\frac{\sqrt{S_\text{BS-}(\omega)}}{L}\right|_\text{LIGO}
   &\simeq 0.98\times 10^{-26}\, \frac{1}{\sqrt\text{Hz}}
\end{align}
\noindent Fig. \ref{fig:GEO600sensitivity} shows the BS Brownian noise in comparison to the other severe limitations of the GEO600 sensitivity, namely the mirror coating Brownian and the beam-splitter thermorefractive noise. Over the whole detection bandwidth of GEO600, the BS Brownian noise is about 35$\%$ lower than the mirror coating Brownian noise. For frequencies larger than 100\,Hz, the BS Brownian noise is more dominant than the BS thermorefractive noise. At the most sensitive frequency of GEO at 500 Hz, the BS noise \eqref{est2GEO} accounts for approximately $10\%$ of the current sensitivity. Fig. \ref{fig:GEO600sensitivity} shows also the sensitivity curve of the proposed GEO-HF upgrade \cite{Grote2010CQG}. This curve so far does not take the BS Brownian noise into account. As illustrated in Fig. \ref{fig:GEO600sensitivity}, the BS Brownian noise impairs the feasible sensitivity of the GEO-HF by about $50\%$ for the frequency range between $50$ and $1000$ Hz.

\section{Conclusion}

In this contribution we applied the direct method of thermal noise calculations from first principles formulated in \cite{18a1TuLeVyPLA}, to the computation of Brownian thermal noise of beam splitters in gravitational wave interferometers. We have demonstrated how the light pressure on both reflective and anti-reflective surfaces contributes to the total thermal noise of the BS. To this end, we took finite sized BS substrates and the coating contributions into account.
An important new ingredient on our calculations is taking into account the stripped pattern of the form-factor that represents the sensitivity of the interferometer’s readout to the BS surface displacement. The pattern is due to the standing waves inside each of the arms\footnote{We note that similar standing-wave patterns are important in computations of thermoelastic \cite{99a1BrGoVy} and thermorefractive \cite{00a1BrGoVy, benthem2009} noise of the BS.} The striped contribution of the pressure turns out to be negligibly small for the substrate Brownian noise. But it provides an increase of the total spectral density of coating Brownian noise by about 50\% (see Sec. IVB of \cite{14a1HeCrGrHiLuNaSiVaVyWi}).

The results show, that the BS noise is negligibly small for advanced LIGO. However, for GEO600, it accounts for about 10\% of the current noise budget in the most sensitive frequency band. Furthermore, BS noise impairs the feasible sensitivity of the proposed GEO-HF design by about 50\%. This is because the additional Fabry-Perot arm cavities in LIGO lead to smaller light powers at the BS, compared to the power circulating in the arms. Hence, the BS noise in comparison to the test mass noise is suppressed. Whereas in GEO600, the contribution of BS noise is much larger, because the light power at the BS is the same as at the test masses. Overall, the Brownian BS noise is not a critical issue for the current sensitivity of gravitational wave detectors. However, the presented approach is based on a first-principle method with the Hamiltonian as starting point. It is thus useful for noise computations in other interferometer topologies, for example Sagnac interferometers or other complex optical devices.

\acknowledgments
We are grateful to Garilynn Billingsley for providing us with the information about the thickness of the coating layers on the aLIGO beamsplitters. 
S.K. acknowledges partial support from EURAMET within project 17FUN05 PhotOQuant. S.V. acknowledges partial support from the Russian Science Foundation (Grant No. 17-12-01095) and from the TAPIR GIFT MSU Support of the California Institute of Technology. This document has LIGO number P1800211-v1.

\appendix

\section{Calculation details for fields and pressures}\label{app1}

In this appendix we present in detail the derivations of the functions $f_\pm$, describing the pressure distributions on both BS sides for two (``plus" and ``minus") modes defined by \eqref{H2}. We consider separately s- and p-polarization of the circulating light waves.

\begin{figure}[b]
\includegraphics[width=0.4\textwidth]{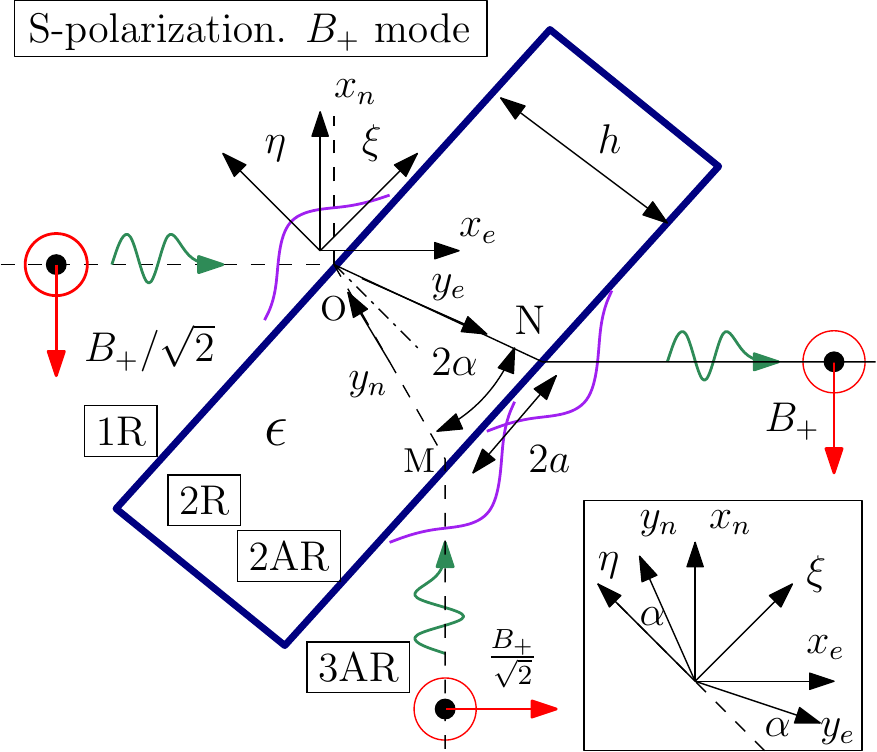}
\caption{$B_+$ mode, s-polarization, i.e. electrical field is normal to the plane of the figure, a fat dot means that it is directed \textit{out of} the plane, a cross \textit{into} the plane, respectively. The red arrows indicate the direction of the magnetic field.}\label{BSplusS}
\end{figure}

\subsection{S-polarization. $B_+$ mode }

For the pressure, we have to calculate the fields on the R and on AR surface inside and outside the BS. We begin with the $B_+$ mode and s-polarization, see Fig.~\ref{BSplusS}, denoted by $B_+$ amplitude of the magnetic field inside the east arm, whereas the wave in north arm is absent. The $z-$axis is directed \textit{out of} plane. The electric field components are denoted as $E$, the magnetic ones by $H$ and subscripts denote projections.

\paragraph*{Surface 1R.}
For the complex amplitudes of electric and magnetic fields we obtain (see Fig.~\ref{BSplusS} and definition \eqref{kbs}):
\begin{subequations}
\begin{align}
 E_z &=\frac{B_+}{\sqrt 2}\left(e^{ikx_e} - e^{-ikx_e}\right)_{\eta=0}\\
     & = i\sqrt 2\, B_+\sin \left(k_{bs} \xi\right), \nonumber \\
 H_n &= -\frac{B_+}{\sqrt 2}\left(e^{ikx_e} + e^{-ikx_e}\right)_{\eta=0}\\
     & = -\sqrt 2\, B_+\cos \left(k_{bs}\xi\right), \nonumber \\
  H_\xi &=  H_\eta= \frac{H_n}{\sqrt 2}=- B_+\cos \left(k_{bs}\xi\right).
\end{align}
\end{subequations}
Using definition \eqref{sigmaD} we calculate the Maxwell stress tensor on surface 1R:
\begin{align}
 \sigma_{\eta\eta} &
 \label{sigma1rSp}
 = - \frac{|B_+|^2}{2\pi}\sin^2 \left(k_{bs}\xi\right).
\end{align}
Component $\sigma_{\eta\eta}$ corresponds to the pressure acting \textit{along} axis $\eta$, hence $p_\eta =\sigma_{\eta\eta}$. The spatial distribution of the pressure $p_\eta$ should be restored in \eqref{sigma1rSp}, by projecting \eqref{f0} on the reflecting surface R:
\begin{align}
  \label{p1rSp}
 p_\eta^{1rS+} &= - \frac{|B_+|^2}{4\pi}\left(1- \cos 2k_{bs}\xi\right)f_\bot^2 ,
\end{align}
see definition \eqref{fbot}.
The pressure consists of a smooth contribution and a fast oscillating one $\sim \cos 2 k_{bs}\xi$.

\paragraph*{Surface 2R (s-polarization, $B_+$ mode)}. The relationship between the electric and magnetic fields inside the material (marked by superscript $\epsilon$) and the light intensity is:
\begin{align}
 |H^\epsilon| & = \sqrt\epsilon\, |E^\epsilon|, \quad I_\epsilon = c\,\frac{\sqrt \epsilon\, |E^\epsilon|^2}{4\pi},\quad 
  n\equiv \sqrt \epsilon.
\end{align}
The intensity inside and outside the BS of AR surface has to be conserved.   
Thus, for the amplitudes of the wave propagating along $y_n-$ and $y_e-$ axes and inside the BS:
\begin{subequations}
\begin{align}
 E_z &= \frac{B_+}{\sqrt n \sqrt 2}\left(e^{inky_n}- e^{_-i nky_n}\right)\\
 &\qquad+ \frac{B_+}{\sqrt n} \left(e^{i nky_e}- e^{_-i nky_e}\right),  \nonumber  \\
  E_z&|_{\eta=0} = \frac{i\sqrt 2\big(\sqrt 2 +1\big) B_+}{\sqrt n}\sin\left( k_\alpha \xi\right),\\
  \label{kalpha}
  k_\alpha & = n k\sin\alpha = k_{bs},\\
  & \sin\alpha  = \frac{1}{\sqrt 2\, n},\quad
      \cos\alpha =\sqrt{1-\frac{1}{2n^2}}\,,\\
  H_\eta &= - \frac{B_+ \sqrt n \sin\alpha}{ \sqrt 2}\left(e^{i nky_n} + e^{_-i nky_n}\right)\\
   &\qquad  - B_+ \sqrt n\sin\alpha \left(e^{i nky_e} + e^{_-i nky_e}\right), \nonumber\\
  H_\eta &|_{\eta=0} = - \sqrt{2 n} \sin \alpha \left(\sqrt 2 + 1\right) B_+\cos k_{bs}\xi,\\
   H_\xi &=  \frac{B_+ \sqrt n \cos\alpha}{ \sqrt 2}\left(e^{i nky_n} + e^{_-i nky_n}\right) \\
     &\qquad - B_+ \sqrt n\cos\alpha \left(e^{i nky_e} + e^{_-i nky_e}\right) \nonumber\\
   H_\xi &|_{\eta=0} =  \sqrt {2n} \cos \alpha \left(1-\sqrt 2 \right) B_+\cos k_{bs}\xi.
\end{align}
\end{subequations}
We calculate the Maxwell stress tensor at the surface 2R, using definition \eqref{sigmaD} again:
\begin{subequations}
\begin{align}
 \sigma_{\eta\eta} 
 \label{sigma2rS+}
  &= - \frac{n |B_+|^2}{4\pi}\left\{3(1+\cos2\alpha) \frac{}{} \right.\\
  &\qquad \left.\frac{}{} 
  -\left[3- 3\cos 2\alpha + 4\sqrt 2 \right] \cos 2k_{bs}\xi\right\}.\nonumber
\end{align}
The spatial distribution of the pressure $p_\eta$ should be restored in \eqref{sigma2rS+}, by projecting \eqref{f0} on the reflecting surface of the BS:
\begin{align}
  \label{p2rSp}
 p_\eta^{2rS+} &= \frac{n |B_+|^2}{4\pi}\left\{3(1+\cos2\alpha) \frac{}{}\right.\\
  &\qquad \left.\frac{}{} 
   -\left[3-3\cos 2\alpha + 4\sqrt 2 \right] \cos 2k_{bs} \xi\right\} f_\bot^2.\nonumber
\end{align}
The total pressure applied to the surface R of BS (s-polarization, $B_+$ mode) is equal to the \textit{sum} of \eqref{p1rSp} 
and \eqref{p2rSp}:
\begin{align}
 p_\eta^{rS+} &= p_\eta^{1rS+} + p_\eta^{2rS+} \\
 \label{prSp}
 & = \frac{ |B_+|^2 f_\bot^2}{4\pi}\left\{3n(1+\cos2\alpha) -1 \frac{}{}\right.\\
   &\qquad \left.
  +\left(1 - n\left[3 - 3\cos 2\alpha + 4\sqrt 2 \right]\right) \cos 2 k_{bs}\xi\right\} ,\nonumber
  \end{align}
We calculate $\mathcal E_+$ of this mode using the assumption \eqref{Lell} and the expression for the function $f_0$ \eqref{f0}:
\begin{align}
   \label{E+}
 \mathcal E_+ &=  \frac{ |B_+|^2 }{\pi}\left(L + \frac{\ell_s +\ell_w}{2}\right)\simeq 
  \frac{ |B_+|^2 L}{\pi}.
\end{align}
Using definitions \eqref{DeltaE}, we get: 
  \begin{align}
  \label{prS+}
 \frac{p_\eta^{rS+}}{\mathcal E_+} &=
    \frac{f_\bot^2}{4L} \left\{3n(1+\cos2\alpha) -1  \frac{}{}\right.\\
   &\qquad \left.
  +\left(1 - n\left[3 - 3\cos 2\alpha + 4\sqrt 2 \right]\right) \cos 2k_{bs}\xi\right\} \nonumber
\end{align}

\end{subequations}

\paragraph*{Surface 3AR (s-polarization, $B_+$ mode).}
The distributions of fields from south and east arm are shifted by $\pm a$ from the center as shown in Fig.~\ref{BSplusS}, $2a= h \tan \alpha$. Using definitions (\ref{fpm}, \ref{xipm}) we obtain:
\begin{subequations}
\begin{align}
E_z &= \frac{B_+}{\sqrt 2}\left(e^{ikx_n} -e^{-ikx_n}\right) f_- \\
    &\qquad + B_+ \left(e^{ikx_e} -e^{-ikx_e}\right) f_+\,,\nonumber\\
E_z&|_{\eta=-h} = 
	i\sqrt 2 B_+ \left(\sin\left(\frac{k\xi_-}{\sqrt 2}\right) f_- \right.\\
	& \qquad \left.+ \sin\left(\frac{k\xi_+}{\sqrt 2}\right) \sqrt 2 f_+\right) , \nonumber\\
  H_\xi &= \frac{B_+}{\sqrt 2}\left(e^{ikx_n} + e^{-ikx_n}\right)\frac{f_-}{\sqrt 2}\\
     & \qquad  - B_+ \left(e^{ikx_e} + e^{-ikx_e}\right)\frac{f_+}{\sqrt 2}\,,\nonumber      \\
   H_\xi &|_{\eta=-h}= 
      B_+ \left( \cos \left(\frac{k\xi_-}{\sqrt 2}\right) f_- \right.\\
   & \qquad \left. - \cos \left(\frac{k\xi_+}{\sqrt 2}\right) \sqrt 2 f_+\right),\quad 
    \\
  H_\eta &= -\frac{B_+}{\sqrt 2}\left(e^{ikx_n} + e^{-ikx_n}\right)\frac{f_-}{\sqrt 2} \\
   &\qquad - B_+ \left(e^{ikx_e} + e^{-ikx_e}\right)\frac{f_+ }{\sqrt 2}\,,\nonumber \\
   H_\eta &|_{\eta=-h}=
    - B_+ \cos \left(\frac{k\xi_-}{\sqrt 2}\right)  f_-  \\
    &\qquad + B_+\cos \left(\frac{k\xi_+}{\sqrt 2}\right)\sqrt 2 f_+ . \nonumber
\end{align}
\end{subequations}
We calculate  Maxwell stress tensor on 3AR surface of BS using definition
\eqref{sigmaD} again:
\begin{align}
 \sigma_{\eta\eta} &=  \frac{|B_+|^2}{4\pi}\left\{-f_-^2 - 2 f_+^2 4\sqrt 2\, f_-f_+ \cos k_{bs}(\xi_++\xi_-)  
  \right.\nonumber \\
  \label{sigma3arS}
    &\qquad \left.+
f_-^2\cos 2\kappa\xi_- + 2f_+^2\cos 2\kappa\xi_+ \right\}.
\end{align}
The pressure along the $\eta$-axis reads:
\begin{align}
 p_\eta^{3arS} &= - \sigma_{\eta\eta}.
\end{align}

\paragraph*{Surface 2AR of the BS (S-polarization, $B_+$ mode): }
\begin{subequations}
\begin{align}
E_z &= \frac{B_+}{\sqrt{2 n}}\left(e^{in ky_n} -e^{-inkx_n}\right) f_- \\
    &\qquad + \frac{B_+}{\sqrt n} \left(e^{inky_e} -e^{-inky_e}\right) f_+\,,\nonumber\\
E_z&|_{\eta=-h} = \frac{i\sqrt 2 B_+}{\sqrt n} \\
    &\qquad \times\left[ f_-\sin k_{bs} \xi_- + \sqrt 2 f_+\sin k_{bs} \xi_+\right],\nonumber\\
 H_\xi &= \frac{B_+\sqrt n}{\sqrt 2}\left(e^{inky_n} + e^{-inky_n}\right)f_-\cos\alpha \\
   &\qquad  - B_+\sqrt n \left(e^{inky_e} + e^{-inky_e}\right)f_+\cos\alpha\,, \nonumber \\
   H_\xi &|_{\eta=-h}= \sqrt 2\sqrt n B_+ \cos\alpha\\
   &\qquad \times    \left( f_-\cos\kappa \xi_- - \sqrt 2 f_+\cos\kappa \xi_+\right),\nonumber\\
  H_\eta &= -\frac{B_+\sqrt n}{\sqrt 2}\left(e^{inky_n} + e^{-inky_n}\right)f_-\sin \alpha \\
   &\qquad + B_+\sqrt n \left(e^{inky_e} + e^{-inky_e}\right)f_+\sin\alpha\,,\nonumber\\
   H_\eta &|_{\eta=-h}= -\sqrt 2\sqrt n\, B_+ \sin \alpha\\
    &\qquad \times\left( f_-\cos\kappa \xi_- + \sqrt 2 f_+ \cos\kappa \xi_+\right). \nonumber
\end{align}
\end{subequations}
Maxwell stress tensor \textit{at} the surface 2AR of the BS:
\begin{align}
 \sigma_{\eta\eta} &=  \frac{n |B_+|^2}{2\pi}
  \left\{-  \left( f_-^2 + 2 f_+^2\right) \cos^2\alpha  \right.\\
  &\qquad  + \left( f_-^2 \cos 2\kappa \xi_- + 2 f_+^2\cos 2 k_{bs}\xi_+\right) \sin^2\alpha \nonumber\\
  &\qquad \left.  + 2\sqrt 2f_+f_-\cos k_{bs}(\xi_++\xi_-)\right\}.\nonumber
\end{align}
Pressure along the $\eta$-axis: $ p_\eta^{2arS+} =  \sigma_{\eta\eta}$.
So, the total pressure along the $\eta-$axis on surface 2AR of the BS (s-polarization, $B_+$ mode) is equal to: 
\begin{align}
 & p_\eta^{arS+}  = p_\eta^{2arS+} +p_\eta^{3arS+},\\
   \label{parS+E}
 & \frac{p_\eta^{2arS+} +p_\eta^{3arS+}}{\mathcal E_+} \\
   &=   \frac{1}{4L}
  \left\{  \left( f_-^2 + 2 f_+^2\right)\left[1 +\frac{1}{n}-2n\right] \frac{}{}\right. \nonumber\\
  &\quad  + 4\sqrt 2(n-1)f_- f_+\cos k_{bs}(\xi_+ + \xi_-)  \nonumber\\
  &\quad \left. +  \left( f_-^2\cos 2k_{bs}\xi_- + 2 f_+^2\cos 2 k_{bs}\xi_+\right)
    \left(-1 +\frac{1}{n}\right) 
    \right\}.\nonumber
\end{align}

\subsection{S-polarization. $B_-$ mode }
\begin{figure}[b]
\includegraphics[width=0.4\textwidth]{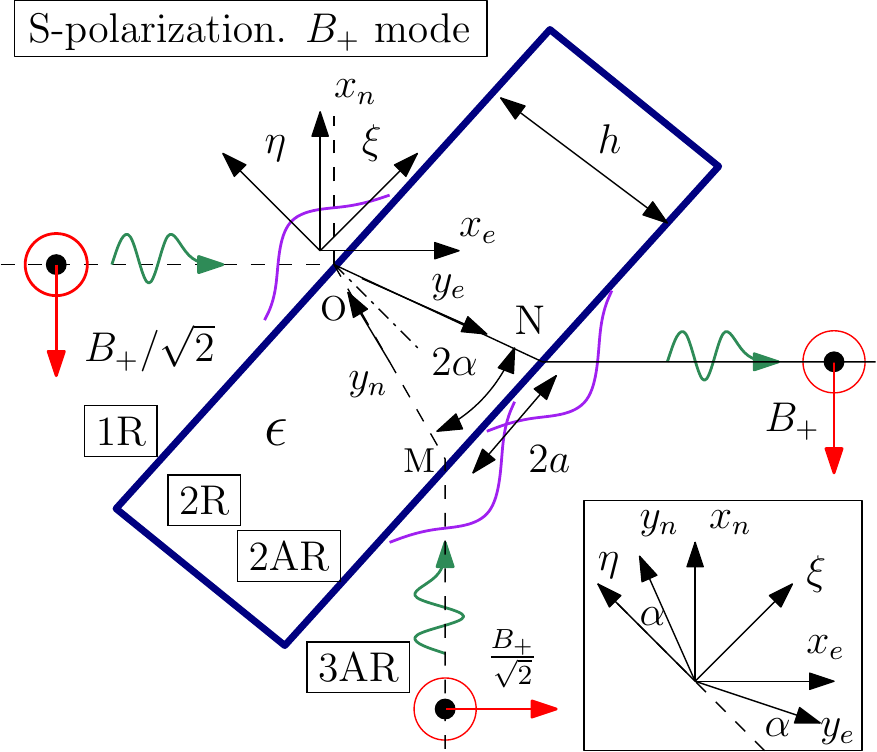}
\caption{$B_-$ mode, s-polarization, i.e. the electrical field is normal to the plane of figure, fat dot means that it is directed \textit{out of} plane, cross means \textit{into} the plane. Red arrows relate to complex amplitudes of magnetic fields.}\label{BSminusS}
\end{figure}

The geometry is shown in Fig.~\ref{BSminusS}, the $z-$axis is directed \textit{out of} the plane of figure. $B_-$ indicates the amplitude of the magnetic field in the north arm, the wave is absent in the east arm.

\paragraph*{Surface 1R of BS (s-polarization, $B_-$ mode):}
\begin{subequations}
\begin{align}
 E_z &=\frac{-B_-}{\sqrt 2}\left(e^{ikx_e} - e^{-ikx_e}\right)
    + B_-\left(e^{ikx_n} - e^{-ikx_n}\right), \\
 E_z&|_{\eta=0}= i \sqrt 2\, B_-\left(\sqrt 2 - 1\right)\sin k_{bs}  ,\\
 H_\eta &|_{\eta=0} = \left( 1-\sqrt 2\right) B_-\cos k_{bs} ) ,\\
  H_\xi &|_{\eta=0}
     = \left( 1+\sqrt 2\right) B_-\cos k_{bs} ) ,
\end{align}
\end{subequations}
Maxwell stress tensor on surface 1R:
\begin{align}
 \sigma_{\eta\eta}
 \label{sigma1rSm}
 &= - \frac{|B_-|^2}{4\pi}\left( 3 +\left[4\sqrt 2 - 3 \right]\cos 2 k_{bs}\xi \right).
\end{align}
The component $\sigma_{\eta\eta}$ corresponds to the pressure acting \textit{along} the $\eta-$axis, hence $p_\eta =\sigma_{\eta\eta}$. A negative sign in \eqref{sigma1rSm} means, that the field \textit{pushes} the BS. The spatial distribution of the pressure $p_\eta$ has to be restored in \eqref{sigma1rSm}, by projecting \eqref{f0} on the reflecting surface of BS:
\begin{align}
  \label{p1rS}
 p_\eta^{1rS-} &= - \frac{|B_-|^2}{4\pi}\left( 3 +\left[4\sqrt 2 - 3 \right]\cos 2k_{bs}\xi\right) f_\bot^2\,,
\end{align}
 $f_\bot$ is defined by \eqref{fbot}.

\paragraph*{Surface 2R of BS (s-polarization, $B_-$ mode):}
\begin{subequations}
 \label{OnRSminus}
\begin{align}
 E_z &=\frac{B_-}{\sqrt 2\sqrt n}\left(e^{ikny_n} - e^{-inky_n}\right),\\  
 E_z &|_{\eta=0}= i \frac{\sqrt 2\, B_-}{\sqrt n}\sin k_{bs}\xi  ,\\
 H_\eta &|_{\eta=0}=  -\sqrt {2n}\,B_- \sin \alpha\, \cos k_{bs}\xi  ,\\
  H_\xi &|_{\eta=0}=\sqrt {2n}\, B_-\cos\alpha \cos k_{bs} \xi ,
\end{align}
\end{subequations}
Maxwell stress tensor on surface 2R:
\begin{align}
 \sigma_{\eta\eta}  
 \label{sigma2rSm}
 &=  \frac{2n |B_-|^2}{4\pi}\left(-\cos^2\alpha+ \sin^2\alpha \cos 2 k_{bs}\xi\right).
\end{align}
Here, the component $\sigma_{\eta\eta}$ corresponds to the pressure acting \textit{along} the $\eta-$axis, hence $p_\eta =-\sigma_{\eta\eta}$. The spatial distribution of the pressure $p_\eta$ has to be restored in \eqref{sigma1rSm}, by projecting \eqref{f0} on the reflecting surface of the BS:
\begin{align}
  \label{p2rS-}
 p_\eta^{2rS-} &= \frac{2n |B_-|^2}{4\pi}\left(\cos^2\alpha - \sin^2\alpha \,\cos 2 k_{bs}\xi\right) f_\bot^2.     
\end{align}
The total pressure acting on surface R (s- polarization, $B_-$ mode) reads:
\begin{align}
\label{prSm}
 p_\eta^{rS-} &= p_\eta^{1rS-} + p_\eta^{2rS-},\\
    \label{prSmb}
    \frac{p_\eta^{rS-}}{\mathcal E_-} &=
      \frac{f_\bot^2}{4 L}\left(2n \cos^2\alpha -3 \right.\\
   &\qquad \left.-\left[4\sqrt 2 -3 +2n \sin^2\alpha \right]\cos 2 k_{bs}\xi\right).\nonumber
\end{align}

\paragraph*{On surface 2AR of the BS (s-polarization, $B_-$ mode):}

Apparently, in this case the fields will be the same, i.e. equation \eqref{OnRSminus} can be applied. Hence, equation \eqref{sigma2rSm} for the stress tensor may be applied, but in this case the component $\sigma_{\eta\eta}$ corresponds to pressure acting \textit{along} the $\eta-$ axis and $p_\eta =\sigma_{\eta\eta}$. The spatial distribution of the pressure $p_\eta$ on the anti-reflecting surface of the BS is equal to:
\begin{align}
  \label{p2arS-}
 p_\eta^{2arS-} &= \frac{2n |B_-|^2}{4\pi}
  \left(-\cos^2\alpha + \sin^2\alpha \,\cos2 k_{bs}\xi_-\right) f_-^2.
\end{align}
Thus, $p_\eta^{2arS-}= - p_\eta^{2rS-}$ (with substitution $f_\bot$ instead of $f_-$).

\paragraph*{Surface 3AR of BS (s-polarization, $B_-$ mode):}
\begin{subequations}
 \label{OnARSminus}
\begin{align}
 E_z &=\frac{B_-}{\sqrt 2}\left(e^{ikx_n} - e^{-ikx_n}\right),\\
   E_z&|_{\eta=0}= i \sqrt 2\, B_- \sin k_{bs}  \xi_- ,\\
 H_\eta &|_{\eta=0} = -B_-  \cos k_{bs}  \xi_- ,\\
  H_\xi &|_{\eta=0} =  B_-\cos k_{bs}  \xi_- ,
\end{align}
\end{subequations}
Maxwell stress tensor on surface 3AR:
\begin{align}
 \sigma_{\eta\eta}  & 
 \label{sigma3arSm}
 = - \frac{ |B_-|^2}{4\pi}\left(1 - \cos 2k_{bs} \xi_-\right)\\
  \label{p3arS-}
 p_\eta^{3arS-} &= \frac{ |B_-|^2}{4\pi}\left(1 - \cos 2k_{bs} \xi_-\right) f_-^2.     
\end{align}
Total pressure acting on surface AR (s-polarization, $B_-$ mode):
\begin{align}
 p_\eta^{arS-} &= p_\eta^{2arS-} + p_\eta^{3arS-},\\
 \label{parS-}
 \frac{p_\eta^{arS-}}{\mathcal E_-}&= \frac{f_-^2}{4 L} \left( 1 +\frac{1}{n} -2n  \right.\\
	& \qquad \left. +\left[\frac{1}{n} -1\right]\cos 2k_{bs}  \xi_-\right).\nonumber
\end{align}

\subsection{P-polarization. $B_+$ mode }

\begin{figure}[b]
\includegraphics[width=0.4\textwidth]{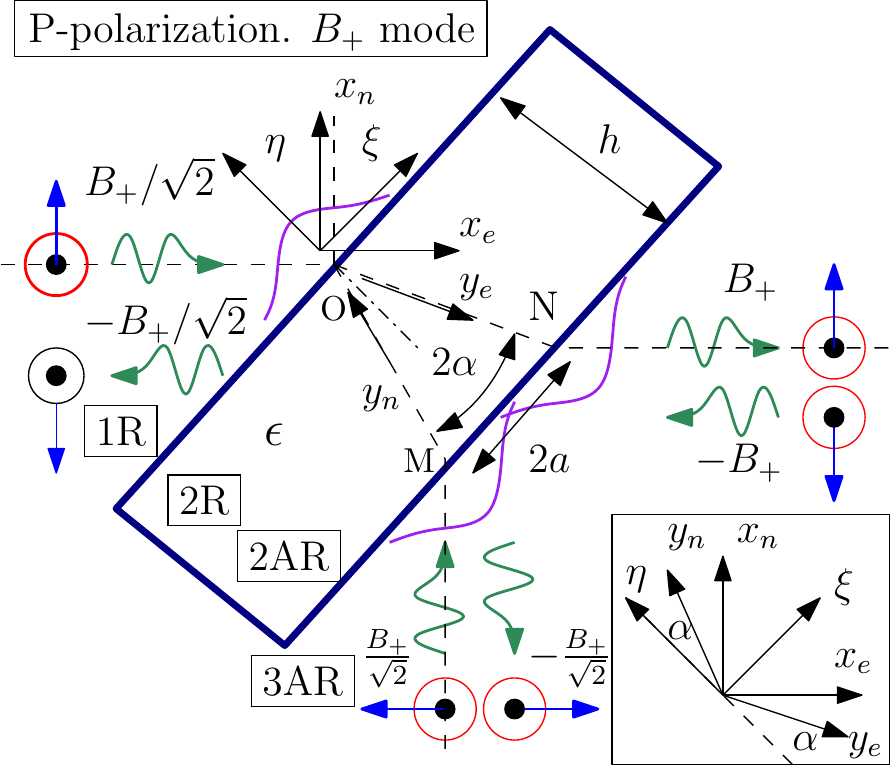}
\caption{$B_+$ mode, p-polarization, i.e. the electric field (blue arrows) is in the plane of figure, the magnetic field is perpendicular to the plane, fat dot means that it is directed \textit{out of} plane, cross means \textit{into} the plane.}\label{BSplusP}
\end{figure}

\paragraph*{On suface 1R of BS (P-polarization, $B_+$ mode):}
\begin{subequations}
\begin{align}
 H_z &=\frac{B_+}{\sqrt 2}\left(e^{ikx_e} + e^{-ikx_e}\right),\\
   H_z &|_{\eta=0} =  \sqrt 2\, B_+\cos k_{bs} \xi,\\
 E_n &= \frac{B_+}{\sqrt 2}\left(e^{ikx_e} - e^{-ikx_e}\right)_{\eta=0} \\
    & = i\sqrt 2\, B_+\sin k_{bs}\xi, \nonumber\\
  E_\xi &=  E_\eta= \frac{E_n}{\sqrt 2}=iB_+\sin k_{bs}\xi.
\end{align}
\end{subequations}
Here, $B_+$ is the amplitude of the electric field in the east arm.
Maxwell stress tensor on surface 1R:
\begin{align}
 \sigma_{\eta\eta} 
 \label{sigma1rP}
   = -\frac{|B_+|^2}{4\pi}\left(1 +\cos\sqrt 2k\xi \right).
\end{align}
Component $\sigma_{\eta\eta}$ corresponds to pressure acting \textit{along} the $\eta-$ axis, hence $p_\eta =\sigma_{\eta\eta}$. 
The spatial distribution of the pressure $p_\eta$ should be restored in \eqref{sigma1rP}, by projecting \eqref{f0} on the reflecting surface of the BS by \eqref{fbot}:
\begin{align}
  \label{p1rP}
 p_\eta^{1rP} &=-\frac{|B_+|^2}{4\pi}\left(1 +\cos\sqrt 2k\xi \right) f_\bot \,.
\end{align}

\paragraph*{On surface 2R (p-polarization, $B_+$ mode):}
\begin{subequations}
\begin{align}
 H_z &= \frac{\sqrt n  B_+}{\sqrt 2}\left(e^{inky_n}+ e^{_-i nky_n}\right) \\
     &\quad + \sqrt n \,B_+ \left(e^{i nky_e} + e^{_-i nky_e}\right),  \nonumber\\
  H_z& |_{\eta=0}  = \sqrt {2n}\big(\sqrt 2 + 1\big) B_+\cos\left( k_{bs} \xi\right),\\
  E_\eta &=  \frac{B_+  \sin\alpha}{\sqrt n \sqrt 2}\left(e^{i nky_n} - e^{_-i nky_n}\right) \\
     &\quad +\frac{B_+}{\sqrt n} \sin\alpha \left(e^{i nky_e} -e^{_-i nky_e}\right), \nonumber\\
  E_\eta& |_{\eta=0} = \frac {i\sqrt 2 B_+ \sin \alpha}{\sqrt n}  \left(\sqrt 2 + 1\right) \sin k_{bs}\xi,\\
   E_\xi &=  -\frac{B_+  \cos\alpha}{ \sqrt n \sqrt 2}\left(e^{i nky_n} - e^{_-i nky_n}\right) \\
     &\quad  + \frac{B_+ \cos\alpha}{\sqrt n} \left(e^{i nky_e} - e^{_-i nky_e}\right),\nonumber\\
   E_\xi &|_{\eta=0} = i\frac{\sqrt 2 \, B_+ \cos \alpha}{\sqrt n} \left(\sqrt 2 - 1\right) B_+\sin k_{bs}\xi\,.
\end{align}
\end{subequations}
Maxwell stress tensor on reflecting surface 2R:
\begin{align}
  \label{sigma2rP}
 \sigma_{\eta\eta}
     & = - \frac{n |B_+|^2}{4\pi}\left\{6\cos^2\alpha  \frac{}{}\right.\\
     &\quad+ \left.\left( 2\left[\sqrt 2 + 1\right]^2 - 6\cos^2\alpha  \right)  \cos 2k_{bs} \xi\right\}.\nonumber
\end{align}
Component $\sigma_{\eta\eta}$ corresponds to pressure acting \textit{along} the $\eta-$ axis , hence $p_\eta =-\sigma_{\eta\eta}$. 

The spatial distribution of pressure $p_\eta$ should be restored in \eqref{sigma2rP}:
\begin{align}
  \label{p2rP}
 p_\eta^{2rP} &=  \frac{n |B_+|^2}{4\pi}\, \left\{6\cos^2\alpha   \frac{}{}\right.\\
     &\quad +\left.
      \left( 2\left[\sqrt 2 + 1\right]^2 - 6\cos^2\alpha 
		\right)  \cos 2k_{bs} \xi\right\}f_\bot^2. \nonumber
\end{align}
Total pressure applied to surface R (p-polarization, $B_+$- mode) is equal to \textit{sum} 
\eqref{p1rP} and \eqref{p2rP}:
\begin{align}
 p_\eta^{rP} &= p_\eta^{1rP} + p_\eta^{2rP}\, ,\\
  \label{FrP+}
  \frac{p_\eta^{rP+}}{\mathcal E_+}&=
  \frac{ f_\bot^2}{4L}
   \left\{6n\cos^2\alpha -1  \frac{}{}\right.\\
     &\quad +\left. \left(2n\left[\sqrt 2 + 1\right]^2 
		  -6n\cos^2\alpha -1\right) \cos 2k_{bs} \xi\right\}.\nonumber
\end{align}

\paragraph*{On surface 3AR (p-polarization, $B_+$ mode).}
Again we have to account that distributions of fields from south arm and east arm are shifted by $\pm a$ \eqref{fpm}
as shown in Fig.~\ref{BSplusP}:
\begin{subequations}
\begin{align}
H_z &= \frac{B_+}{\sqrt 2}\left(e^{ikx_n} + e^{-ikx_n}\right) f_- \\
    &\qquad  + B_+ \left(e^{ikx_e} + e^{-ikx_e}\right) f_+\,, \nonumber\\
H_z&|_{\eta=-h}  = \sqrt 2 B_+ \left(f_- \cos k_{bs}\xi_-)  +  \sqrt 2 f_+\cos k_{bs}\xi_+\right),  \\
 E_\xi &= - \frac{B_+}{\sqrt 2}\left(e^{ikx_n} - e^{-ikx_n}\right)\frac{f_-}{\sqrt 2} \\
   &\qquad  + B_+ \left(e^{ikx_e} - e^{-ikx_e}\right)\frac{f_+}{\sqrt 2}\,,\nonumber\\
   E_\xi & |_{\eta=-h}= i B_+ \left( \sqrt 2 f_+ \sin k_{bs}\xi_+- f_-\sin k_{bs}\xi_-\right),\\
  E_\eta &= \frac{B_+}{\sqrt 2}\left(e^{ikx_n} - e^{-ikx_n}\right)\frac{f_-}{\sqrt 2} \\
	 &\qquad + B_+ \left(e^{ikx_e} - e^{-ikx_e}\right)\frac{f_+ }{\sqrt 2}\,,\nonumber\\
   E_\eta & |_{\eta=-h}= i B_+ \left( f_-\sin k_{bs}\xi_-+ \sqrt 2 f_+\sin k_{bs}\xi_+\right).
\end{align}
\end{subequations}
Maxwell stress tensor on anti-reflecting surface 3AR: 
\begin{align}
 \label{sigma3arP}
 \sigma_{\eta\eta} &= - \frac{|B_+|^2}{4\pi}\left\{ f_-^2 +2 f_+^2  \right.\\
      &\quad  +2\sqrt 2 f_-f_+ \cos k_{bs}\left(\xi_++ \xi_-\right)\nonumber\\
		&\qquad \left. + \left[f_-\cos 2k_{bs}\xi_- + \sqrt 2 f_+\cos 2k_{bs}\xi_+\right]^2 \right\}.\nonumber
\end{align}
Pressure along $\eta-$axis :
\begin{align}
 \label{p3arP}
 p_\eta^{3arP} &= - \sigma_{\eta\eta}.
\end{align}

\paragraph*{On surface 2AR of the BS (p-polarization, $B_+$ mode):}
\begin{subequations}
\begin{align}
H_z &= \frac{\sqrt n\,B_+}{ \sqrt 2}\left(e^{in ky_n} +e^{-inkx_n}\right) f_- \\
    &\qquad + \sqrt n\,B_+ \left(e^{inky_e} + e^{-inky_e}\right) f_+\,,\nonumber\\
 H_z&|_{\eta=-h} = \sqrt{ 2 n}\, B_+ 
    \left( f_- \cos k_{bs}\xi_-+ \sqrt 2 f_+ \cos k_{bs}\xi_+\right),\\
 E_\xi &= -\frac{B_+}{\sqrt {2 n}}\left(e^{inky_n} - e^{-inky_n}\right)f_-\cos\alpha \\
    &\qquad  + \frac{B_+}{\sqrt n} \left(e^{inky_e} - e^{-inky_e}\right)f_+\cos\alpha\,,\nonumber\\
 E_\xi &|_{\eta=-h}= i \frac{\sqrt 2 B_+}{\sqrt n}\cos\alpha\\
 &\qquad \times
    \left(  \sqrt 2 f_+\sin k_{bs}\xi_+ -f_-\sin k_{bs}\xi_-\right),\nonumber\\
 E_\eta &= \frac{B_+}{\sqrt {2n}}\left(e^{inky_n} - e^{-inky_n}\right)f_-\sin \alpha \\
   &\qquad+ \frac{B_+}{\sqrt n} \left(e^{inky_e} - e^{-inky_e}\right)f_+\sin\alpha\,,\nonumber\\
 E_\eta &|_{\eta=-h}= \frac{\sqrt 2\, B_+}{\sqrt n} i\sin \alpha
    \left( f_-\sin k_{bs}\xi_- + \sqrt 2 f_+\sin k_{bs}\xi_+\right).
\end{align}
\end{subequations}

\noindent Maxwell stress tensor on surface 2AR
 \begin{align}
  \label{sigma2arP}
  \sigma_{\eta\eta}= & \frac{n |B_+|^2}{2\pi}
  \left\{- \left( f_-^2  + 2 f_+^2\right) \cos^2\alpha \right.\\
   &\quad +\left[-\left( f_-^2\cos 2k_{bs}\xi_- + 2 f_+^2\cos 2k_{bs}\xi_+\right) \sin^2\alpha \right.\nonumber\\
   &\qquad \left.\left. - 2\sqrt 2 f_-f_+\cos k_{bs}(\xi_-+\xi_+)\right] \right\}. \nonumber
 \end{align}
Pressure along $\eta-$axis:
\begin{align}
 p_\eta^{2arP+} &=  \sigma_{\eta\eta}.
\end{align}
Total pressure along the $\eta-$ axis acting on surface AR (p-polarization, $B_+$ mode) is equal to:
\begin{align}
  p_\eta^{arP+} & = p_\eta^{2arP+} +p_\eta^{3arP+} \\
   \label{parP+}
 \frac{p_\eta^{arP+}}{\mathcal E_+}&= 
  \frac{ |B_+|^2}{4\pi}
  \left\{  \left( f_-^2 + 2 f_+^2\right)\left[1 +\frac 1 n -2n\right] \frac{}{}\right.\\
     & +  \left(4(1-n)\sqrt 2f_- f_+ \cos k_{bs}(\xi_- +\xi_+)\right.\nonumber\\
      +& \left. \left. \left( f_-^2\cos 2k_{bs}\xi_- + 2 f_+^2\cos 2k_{bs}\xi_+\right)
    \left(1- \frac 1 n \right) \right) \frac{}{} \right\}.\nonumber
\end{align}

\subsection{P-polarization. $B_-$ mode }

\begin{figure}[b]
\includegraphics[width=0.4\textwidth]{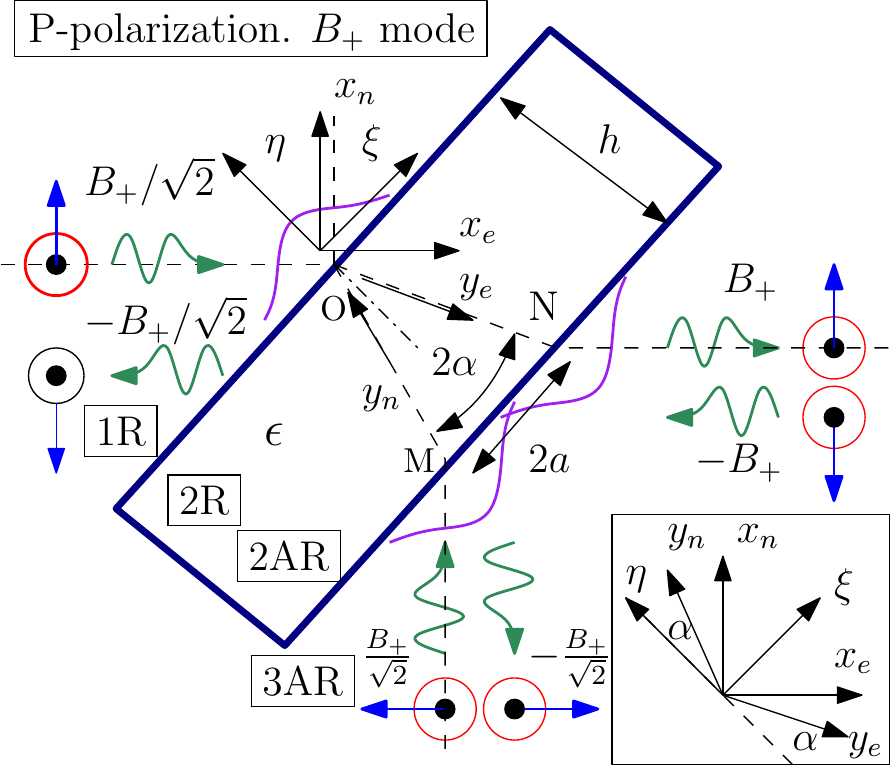}
\caption{$B_-$ mode, p-polarization, i.e. the magnetic field is normal to the plane of figure, fat dot means that it is directed 
\textit{into} the plane, cross means \textit{out of} plane. $B_-$ is the amplitude of the electric field in the north arm. 
Blue arrows indicate the direction of the electric field.}\label{BSminusP}
\end{figure}

\paragraph*{On surface 1R (p-polarization, $B_-$ mode): }
\begin{subequations}
\begin{align}
 H_z &=\frac{-B_-}{\sqrt 2}\left(e^{ikx_e} + e^{-ikx_e}\right)\\
     &\quad + B_-\left(e^{ikx_n} + e^{-ikx_n}\right), \nonumber\\
 H_z &|_{\eta=0}=  \sqrt 2\, B_-\left(\sqrt 2 - 1\right)\cos  k_{bs}\xi ,\\
 E_\eta &|_{\eta=0} = i\left( \sqrt 2 -1\right) B_-\sin  k_{bs}\xi ,\\
  E_\xi &|_{\eta=0} = -i\left( 1+\sqrt 2\right) B_-\sin  k_{bs}\xi ,
\end{align}
\end{subequations}
Maxwell stress tensor on surface 1R:
\begin{align}
   \label{sigma1rPm}
 \sigma_{\eta\eta} &= - \frac{|B_-|^2}{4\pi}\left( 3 +\left[3 - 4\sqrt 2 \right]\cos  2 k_{bs}\xi\right).
\end{align}
Component $\sigma_{\eta\eta}$ corresponds to the pressure acting \textit{along} the $\eta-$axis, hence $p_\eta =\sigma_{\eta\eta}$.  The spatial distribution of the pressure $p_\eta$ should be restored in \eqref{sigma1rPm}, by projecting \eqref{f0} on surface 1R:
\begin{align}
  \label{p1rP-}
 p_\eta^{1rP-} &= - \frac{|B_-|^2}{4\pi}\\
 &\qquad \times\left( 3 +\left[3 -4\sqrt 2 \right]\cos  2 k_{bs}\xi\right) f_\bot^2\,.\nonumber
\end{align}

\paragraph*{On surface 2R (s-polarization, $B_-$ mode):}
\begin{subequations}
 \label{OnRPminus}
\begin{align}
 H_z &=\frac{\sqrt n \,B_-}{\sqrt 2}\left(e^{iky_n} + e^{-iky_n}\right),\\    
      H_z &|_{\eta=0}=  \sqrt {2n}\, B_- \cos  k_{bs}\xi ,\\
 E_\eta &|_{\eta=0} =  i\frac{\sqrt {2}\,B_-}{\sqrt n} \sin \alpha\, \sin  k_{bs}\xi ,\\
 E_\xi &|_{\eta=0} =-i \frac{\sqrt {2}\, B_-}{\sqrt n}\cos\alpha \sin  k_{bs}\xi ,
\end{align}
\end{subequations}
Maxwell stress tensor on surface 2R:
\begin{align}
 \label{sigma2rPm}
 \sigma_{\eta\eta} 
 &=  \frac{2n |B_-|^2}{4\pi}\left(-\cos^2\alpha - \sin^2\alpha \cos 2 k_{bs}\xi\right).
\end{align}
Here component $\sigma_{\eta\eta}$ corresponds to pressure acting \textit{along} the $\eta-$axis, hence $p_\eta =-\sigma_{\eta\eta}$.  The spatial distribution of the pressure $p_\eta$ should be restored in \eqref{sigma2rPm}, by projecting \eqref{f0} on the reflecting surface of the BS:
\begin{align}
  \label{p2rP-}
 p_\eta^{2rP-} &= \frac{2n |B_-|^2}{4\pi}\\
 &\qquad \times \left(\cos^2\alpha + \sin^2\alpha \,\cos 2 k_{bs}\xi\right) f_\bot^2.   \nonumber  
\end{align}
Total pressure acting on surface R (p- polarization, $B_-$ mode):
\begin{align}
 p_\eta^{rP-} &= p_\eta^{1rP-} + p_\eta^{2rP-},\\
 \label{prPm}
 \frac{p_\eta^{rP-}}{\mathcal E_-}& =
     \frac{f_\bot^2}{4L}\left(2n \cos^2\alpha -3 \right.\\
     &\qquad \left. + \left[4\sqrt 2 -3 +2n \sin^2\alpha \right]\cos 2 k_{bs}\xi\right) .\nonumber
\end{align}

\paragraph*{On surface 2AR (p-polarization, $B_-$ mode).}
In this case equation \eqref{sigma2rPm} for the stress tensor may be applied, but in this case component $\sigma_{\eta\eta}$ corresponds to pressure acting \textit{along} the $\eta-$axis and $p_\eta 
=\sigma_{\eta\eta}$. The spatial distribution of the pressure $p_\eta$ on the anti-reflecting surface of the BS is equal to:
\begin{align}
  \label{p2arP-}
 p_\eta^{2arP-} &= - \frac{2n |B_-|^2}{4\pi}\\
  &\qquad \times\left(\cos^2\alpha + \sin^2\alpha \,\cos\sqrt 2 k\xi_-\right) f_-^2. \nonumber    
\end{align}
Obviously, $p_\eta^{2arP-}= - p_\eta^{2rP-}$.

\paragraph*{On surface 3AR (p-polarization, $B_-$ mode):}
\begin{subequations}
 \label{OnARPminus}
\begin{align}
 H_z &=\frac{B_-}{\sqrt 2}\left(e^{ikx_n} + e^{-iky_n}\right),\\  
   H_z &|_{\eta=0}=  \sqrt 2\, B_- \cos k_{bs} \xi_- ,\\
 E_\eta &|_{\eta=0}=  iB_-  \sin k_{bs} \xi_- ,\\
 E_\xi &|_{\eta=0}= -i B_-\sin k_{bs} \xi_- .
\end{align}
\end{subequations}
Maxwell stress tensor on surface 3AR:
\begin{align}
 \sigma_{\eta\eta}  &
 \label{sigma3arPm}
 = - \frac{ |B_-|^2}{4\pi}\left(1 + \cos 2 k_{bs}\xi_-\right)\\
  \label{p3arP-}
 p_\eta^{3arP-} &= \frac{ |B_-|^2}{4\pi}\left(1 + \cos 2 k_{bs}\xi_-\right) f_-^2.     
\end{align}

\paragraph*{Total pressure acting on surface AR (p- polarization, $B_-$ mode):}
\begin{align}
 p_\eta^{arP-} &= p_\eta^{2arP-} + p_\eta^{3arP-}\\
   \label{parP-}
 \frac{p_\eta^{arP-}}{\mathcal E_-}&=
      \frac{f_-^2}{4 L} \left( 1 -2n \cos^2\alpha \frac{}{}\right.\\
    &\qquad \left.	+\left[1 -2n \sin^2\alpha \right]\cos 2 k_{bs}\xi_-\frac{}{}\right).\nonumber
\end{align}

\end{document}